**Characterization of vertically aligned carbon nanotube forests grown on stainless steel surfaces**

*Eleftheria Roumeli[1,2], Marianna Diamantopoulou[1], Marc Serra-Garcia[1,3], Paul Johanns[4], Giulio Parcianello[5], Chiara Daraio[2*]*

[1] Department of Mechanical and Process Engineering, Swiss Federal Institute of Technology (ETH Zurich), 8092, Zurich, Switzerland.

[2] Division of Engineering and Applied Science, California Institute of Technology, Pasadena, CA 91125, USA.

[3] Department of Physics, Swiss Federal Institute of Technology (ETH Zurich), 8092, Zurich, Switzerland.

[4] Institute of Mechanical Engineering, Ecole Polytechnique Fédérale de Lausanne, 1015 Lausanne, Switzerland.

[5] General Electric Switzerland, CH-5401 Baden, Switzerland

Vertically aligned carbon nanotube (CNT) forests are a particularly interesting class of nanomaterials, because they combine multifunctional properties, such as high energy absorption, compressive strength, recoverability and super-hydrophobicity with light weight. These characteristics make them suitable for application as coating, protective layers and antifouling substrates for metallic pipelines and blades. Direct growth of CNT forests on metals offers the possibility to transfer the tunable CNT functionalities directly onto the desired substrates. Here, we focus on characterizing the structure and mechanical properties, as well as wettability and adhesion of CNT forests grown on different types of stainless steel. We investigate the correlations between composition and morphology of the steel substrates with the micro-structure of the CNTs, and reveal how the latter ultimately controls the mechanical and wetting properties of the CNT forest. Additionally, we study the influence of substrate morphology on the adhesion of CNTs to their substrate. We highlight that the same structure-property relationships govern the mechanical performance of CNT forests grown on steels and on Si.

1. Introduction

The unique properties of carbon nanotube (CNT) forests have been documented extensively in the past decade [1–4]. Modifying the synthesis conditions allows tuning of the geometry, alignment, density and structure of CNTs, which offers control over their mechanical performance [5–10]. CNT forests grown directly on metal substrates, and specifically on steels, are particularly interesting because they offer a conformal coating solution, which is independent from the substrate's geometry [11–36]. Direct growth avoids the need for intermediate adhesive layers, thus enabling a robust contact interface between the CNTs and the metal [12].

Floating catalyst chemical vapor deposition combines high CNT yield, vertical alignment and conformal substrate coverage. This method is commonly used for CNT forest growth on metal substrates [12,28,29,31]. Most literature on CNTs grown on metallic substrates, either in the form of aligned forests or non-aligned CNTs, focuses on the synthesis methods [11,15,18,20,26,29,31,34–37]. Few reports characterize the CNTs super-hydrophobicity [21,27,28], corrosion resistance [27,38], field-emission properties [14,25] and electrochemical performance as capacitors [12,13,16]. The mechanical response of CNT forests grown on metallic substrates, like their behavior under compression or impact, as well as their adhesion to the substrate, have not yet been thoroughly investigated.

The remarkable compressive behavior of CNT forests grown on silicon has been documented in literature and it includes energy dissipation in the order of MJ/m[36,39,40], compressive strengths ranging between 0.1-500 MPa [9,39] and almost full-length strain recovery (>80%) for hundreds of compression cycles [1,41]. The mechanical performance of the silicon-grown CNT forests has been correlated to the forests' structure and alignment [6,8,9,41]. The forest's structural properties, in turn, depend on the catalyst distribution and interfacial effects between the CNTs and the silicon substrates [9,42,43]. However, the behavior of catalysts at the interfaces with the steel substrates is different, due to the native composition of these substrates. Steels, for example, are rich of multiple metals and oxides. In the floating catalyst method, the interactions of the supplied metal catalyst particles with the native metals and oxides of the growth substrate determine the rate and yield of CNT formation [31]. We expect the differences in surface composition between silicon and steel substrates, to affect the morphology and ultimate mechanical behavior of the CNT forests grown in the different types of substrates. The focus of this work is to study the mechanical properties, adhesion and wettability of CNT forests grown on steel substrates and reveal the structure-property relationships that govern the CNT performance. Our results demonstrate the mechanical performance of CNT forests grown on steel is governed by the CNT micro-structure on the same way as it does on Si-grown CNTs.

Considering the fact that the CNT forests combine light-weight, super-hydrophobicity, high energy absorption, similar to that of rigid polymer foams, with reusability, due to the large strain recoverability, a wide area of steel coating applications can be reached.

2. Experimental
  2.1. Materials
    2.1.1. Steel substrates

The compositions of the used steels are summarized in Table 1. Prior to growth, all the substrates were placed in a Cr-etching bath for 15 min and then ultra-sonicated in acetone for 5 min to remove any organic residues or dust. In addition, the substrates were air-annealed for 30 min at 827ºC before the start of the growth process (section 2.1.2). The surface roughness of all substrates was measured with a Dektak XT Stylus profilometer and the values are summarized in Table 1.

The steels' surface energy was determined through contact angle measurements using a Drop Shape Analyzer from Krüss Optronic GmbH, model DSA25[44]. Young's theory relates the measured contact angle ($\theta$) to the specific free energy of the tested solid ($\gamma_S$), the used liquid ($\gamma_L$) and the interfacial free energy between the solid and the liquid ($\gamma_{SL}$) as follows[44]:

$$\gamma_L cos\theta = \gamma_S - \gamma_{SL} \qquad \text{Eq. (1)}$$

Fowkes' theory predicts the surface energy of a solid, assuming it is a summation of individual and independent components (polar, dispersive, acidic etc.):

$$\gamma_S = \gamma_s^p + \gamma_s^d + \gamma_s^{ab} + \cdots \quad \text{Eq. (2)}$$

where the superscripts "*d*", "*p*" and "*ab*" correspond to the dispersive, polar and acid-base interactions respectively.

We used water and diiodomethane to probe the polar and dispersive forces of the substrates and a harmonic-mean approach to sum their contribution following Wu's equation[44]:

$$(1 + cos\theta)\gamma_L = 4\left(\frac{\gamma_L^d \gamma_S^d}{\gamma_L^d + \gamma_S^d} + \frac{\gamma_L^p \gamma_S^p}{\gamma_L^p + \gamma_S^p}\right) \quad \text{Eq. (3)}$$

where the subscript "*L*" is for each liquid used in the measurement

**Table 1.** Stainless steel substrates nominal compositions and surface characteristics

| Steel sample | Cr (wt%) | Ni (wt%) | Mo (wt%) | Cu (wt%) | C (wt%) | Surface roughness (μm) | Surface energy, $\gamma_S$ (mJ/m²) |
|---|---|---|---|---|---|---|---|
| S1 | 1.9-2.6 | | | <0.3 | 0.05-0.15 | 0.2 | 35.4 |
| S2 | 8-9.5 | <0.4 | | | 0.08-0.12 | 0.13 | 30.7 |
| S3 | 11-12.5 | 2.2-3 | 1.3-1.9 | | 0.1-0.14 | 0.36 | 53.7 |
| S4 | 13.5-16 | 4-6 | 1.2-2 | 1-2 | 0.07 | 0.85 | 44.8 |

### 2.1.2. CNT forest synthesis

CNT forests were synthesized on thermally oxidized silicon (Si) wafers and/or on the different stainless steel substrates, using vapor phase or "floating catalyst" thermal CVD. The furnace tube had an external diameter of 5 cm and a 15 cm heating zone [10]. The synthesis reactions were performed in Ar flow of 800 sccm at 827 °C in atmospheric pressure. The precursor solution was ferrocene in toluene, at a concentration of 0.02 g/ml and the injection rate was 1 ml/min.

### 2.2. Characterization methods

To characterize the CNT structure and purity we used scanning electron microscopy (SEM) and Raman spectroscopy. SEM images were obtained using a HITACHI, SU8200 SEM, operating at 2-5 kV and 11nA (Fig. 1). High resolution transmission electron microscopy (HRTEM) images were obtained using a FEI TF30ST operating at 300 kV. Raman spectroscopy measurements were performed on a NTEGRA Spectra, NT-MDT. The hydrophobicity response of the CNT forest was determined through the water contact angle measurements described in the previous section.

The micromechanical testing of the samples was performed using a FT-MTA02 module from FEMTO TOOLS. The obtained data of this experiment include also contributions from the sensor and system configuration as well as from the supporting steel material underneath the CNT forest. In order to remove these contributions and allow further calculations on the response of the CNT forests, we tested also a polished uncoated SS substrate[45]. At least four experiments for every sample type were performed, and the average values for dissipated energy, unloading modulus (*E*) and peak stress are extracted from the obtained stress-strain curves

The samples' recovery, *R*, is defined as the displacement recovered at the end of each compression cycle, divided by the maximum displacement, according to the following formula[10]:

$$R = \frac{\varepsilon_{max} - \varepsilon_{unload}}{\varepsilon_{max}}. \quad \text{Eq. (4)}$$

Here, $\varepsilon_{max}$ is the maximum displacement at the end of compression cycle, and $\varepsilon_{unload}$ is the displacement after unloading to 10% of the maximum load in each cycle.

The loss coefficient, *η*, measures energy dissipation and is calculated as follows[46]:

$$\eta = \frac{\Delta U_i}{2\pi U_1}, \quad \text{Eq. (5)}$$

where $U_1$ is the elastic energy stored in the material when it is loaded elastically to a stress $\sigma_{max}$ during the first cycle and $\Delta U_i$ is the energy dissipated in the $i^{th}$ cycle.

Several methods have been reported to study the CNT adhesion to metal substrates including peel tests, scotch tape tests, scratch tests, indentation and combinations of those [47,48]. Amongst those, the scratch test is the most widely accepted method to quantitatively assess well-adherent coatings on metals[47,49,50]. To study specifically the adhesion of CNTs to their growth substrates, the literature reported methods include: the scotch tape removal method [27,28], the pull-off method, which involves taping over the CNTs and measuring the pulling force[12,30], and the ultrasonication method, in which the time required to remove the CNTs from the substrate is recorded[27,33]. However, these methods are often user-sensitive, uncalibrated and do not guarantee that the CNTs will be removed from the substrate as opposed to breaking at weak points[51]. More accurate methods proposed for adhesion testing employ a nano-indenter or an atomic force microscope (AFM)[51,52], which scratches through the substrates and measures the coating adhesion strength. Nevertheless, this approach is bound by a low force detection threshold and the dimensions of the coating may impose additional restrictions. Thus, in order to study the adhesion of CNT forests to their growth substrates, we develop a custom-built scratch setup.

The custom setup consists of a 3-axis linear stage holding a razor blade (VWR International, No. 9) adjusted at a 45° angle above the substrate. The substrate is stabilized with four screws. A strain gauge is placed in the x-direction (scratching direction), to detect the force on the substrate while the razor blade scratches off the CNTs forest. After positioning the sample on the stage using the adjustment screws, the razor blade is moved manually, to get in contact with the edge of CNT forests. The samples are scratched with five different scratching speeds, namely 0.2, 0.5, 1.0, 2.0 and 5.0 mm/s. The output voltage of the strain gauge is recorded as a function of time with an oscilloscope (Tektronix, DP0314, Digital Phosphor Oscilloscope). After calibration ($K_c$=0.045 V/N), the scratching force can be obtained. From the force, the energy release per unit area (E/A, in J/m²) can be calculated as follows:

$$\frac{E}{A} = \frac{E}{w \cdot L} = \frac{E}{w \cdot u \cdot t} = \frac{P}{w \cdot u} = \frac{F \cdot u}{w \cdot u} = \frac{F}{w}. \quad \text{Eq. (6)}$$

Here, *w* is the width of the CNT forest, *P* is the power in J/s, *L* is the length of the forest and *u* is the velocity in m/s.

3. Results and Discussion

   3.1. Structural Investigation

In Fig. 1, we present typical SEM and HRTEM images of CNTs forests grown on the S1 and S4 substrates. We observe a conformal coating of about 60-80 μm in every case on the top of the steel surface. The average growth rate is the same across all samples grown on steels, 1.4±0.2 μm/min. The average outer nanotube diameter, as calculated from SEM images, is 93 ± 20 nm for CNTs grown on S1-S2 and 65 ± 20 nm for CNTs grown on S3-S4. HRTEM images reveal that CNTs grown on S1-S2 have on average 107 walls, while those on S3-S4 have on average 68 walls. The small height of the grown CNT forests does not allow accurate direct mass measurement. To calculate their mass, we employ the method suggested by Laurent et al [53]. In it, the mass and volume of a single multi-walled CNT, $m_{MWCNT}$ and volume $V_{MWCNT}$, can be calculated as follows:

$$m_{\text{MWCNT}} = \frac{1}{A}\pi L[nd_{out} - 2d_{s-s}\sum_{i=0}^{n} i] \quad \text{(Eq.7)}$$

$$V_{MWCNT} = \frac{\pi L d_{out}^2}{4} \quad \text{(Eq. 8)}$$

Here, $A$ is the specific surface area of a single nanotube sheet (equivalent to the one side of a rolled graphene sheet regardless of its diameter, $A$=1315 m$^2$/g), $L$ is the CNT length, $n$ is the number of walls of the CNT, $d_{out}$ is the outer diameter, $d_{s-s}$ is the inter-tube distance ($d_{s-s}$ = 0.34 nm). Then from the SEM-observed areal density (7.9x10$^8$ CNTs/cm$^2$ for S1-S2 and 1.4x10$^9$ CNTs/cm$^2$ for S3-S4) and the overall forest volume (area of steel substrate x forest height), we calculate the sample density. Based on HRTEM and SEM observations, this method results on a density of 0.1 g/cm$^3$ for samples S1-S2, and 0.06 g/cm$^3$ for S3-S4, which are within the literature-reported values for CNT forests [6,10,54].

Differences on surface properties and composition of the steel substrates can explain the differences in CNT diameters and density for the various substrates. As can be seen from Table 1, steels S3-S4 are richer in iron alloys like Fe-Ni and Fe-Cr, compared to S1-S2. In literature, excess of Cr and Ni over Fe, has been associated with the formation of CNTs with smaller diameter and narrower distribution which is in agreement with our observations of CNTs with smaller diameters in samples S3-S4 [55,56]. Additionally, samples S3 and S4 have a higher surface roughness and a higher surface energy compared to S1 and S2. The higher surface roughness combined with better wettability can inhibit catalyst particle mobility and thus, hinter particle coalescence [57,58]. Therefore, on S3-S4 the available catalyst size would be smaller thus leading to smaller CNT diameters, which is the result of our experimental observation [58,59]. The larger CNT diameter and broader distribution found on samples S1-S2 can be justified by the significantly lower Cr and Ni content, as well as the smoother surface and lower surface energy of these steels. Lower surface roughness and energy lead to catalyst particles with higher mobility, which would allow larger clustering and thus, the SEM-observed CNTs with larger diameters can be justified.

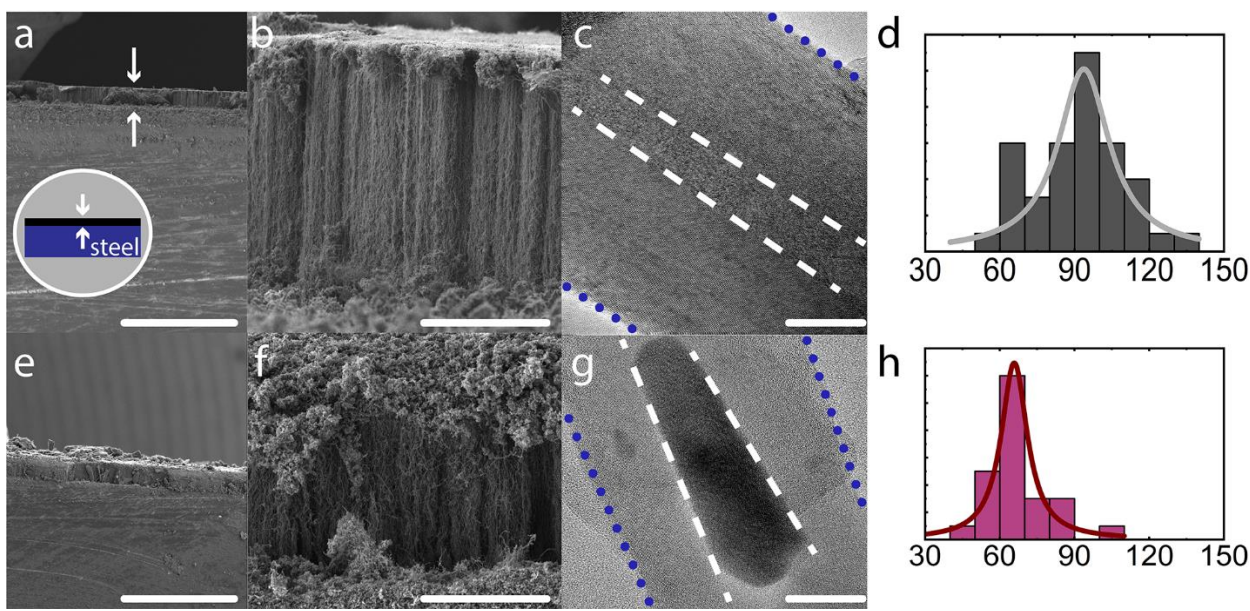

**Figure 1.** (a-b) Representative SEM images of sample S1; (c) HRTEM image of individual CNT from sample S1; dash lines mark the inner tube wall and dotted lines mark the outer CNT wall; (d) outer diameter size distribution for sample S1; (e-f) SEM images of sample S4; (g) HRTEM image of individual CNT from sample S4; (h) measured outer diameter size distribution for sample S4. Scale bars correspond to: (a, e) 500 µm, (b, f) 40 µm, (c, g) 10 nm.

To evaluate the possible differences between CNT forests grown on Si wafers versus steel substrates, we perform control experiments. From SEM images of CNTs grown on silicon, presented in Supplementary Material (Fig. S.1a-b), the measured forest height is 1.0 ± 0.05 mm while the CNT diameter is 70 ± 10 nm. The average density is 0.21 ± 0.08 g/cm$^3$ and growth rate is 20 µm/min. These features are in agreement with previous reports on CNT forests grown via the floating-catalyst or fixed catalyst processes [40,60,61]. The observed average diameter of Si-grown CNTs is between the diameters of CNTs grown on steels. The higher growth rate and density of the forests grown on Si, can be justified by the substrate composition. In the case of Si wafer substrate, ferrocene feedstock is the only source of catalyst particles, whereas on the steel substrates, since the surface is rich in metals such as Fe, Cr, Ni, Mo and Cu, the corresponding metal oxides are formed during annealing at elevated temperatures in air flow, serving as additional catalyst sites[11,17,30]. In Si substrates the dissociated carbon atoms are diffusing over Fe catalyst particles, while on steel substrates there are additional sites available composed primarily of Cr and Ni. Carbon solubility in Cr and Ni is lower compared to Fe [55,62]. As such, the lower growth rate, and subsequently the lower density, in steel substrates can be justified by the presence of Cr and Ni, in agreement with previous reports from Chen et al [55].

To analyze the quality of the produced CNTs, we perform Raman spectroscopy measurements (Fig. 2). The Raman peak at around 1600 cm$^{-1}$ (G band) is associated to the graphitic carbon, while the one around 1340 cm$^{-1}$ (D band) is related to a hybridized vibrational mode arising from defects on the graphene structure. The ratio between G and D peaks ($I_{G/D}$) can be used as an indication of the CNTs' quality[51]. CNT forests grown on Si wafers result in a high $I_{G/D}$ ratio of 1.9 indicating their high quality, while the CNTs grown on the steel substrates range from 0.7-1.0. The better alignment and purity of the CNTs grown on Si versus steels as suggested by Raman, is in agreement with the previous SEM observations. The small variation between samples grown on S1-S2 and S3-S4, with the former presenting higher values (0.9-1.0) compared to the latter (0.71-0.79), suggest a lower defect density and higher degree of alignment for forests grown on S1-S2 in comparison to those grown on S3-S4 steels [51,63].

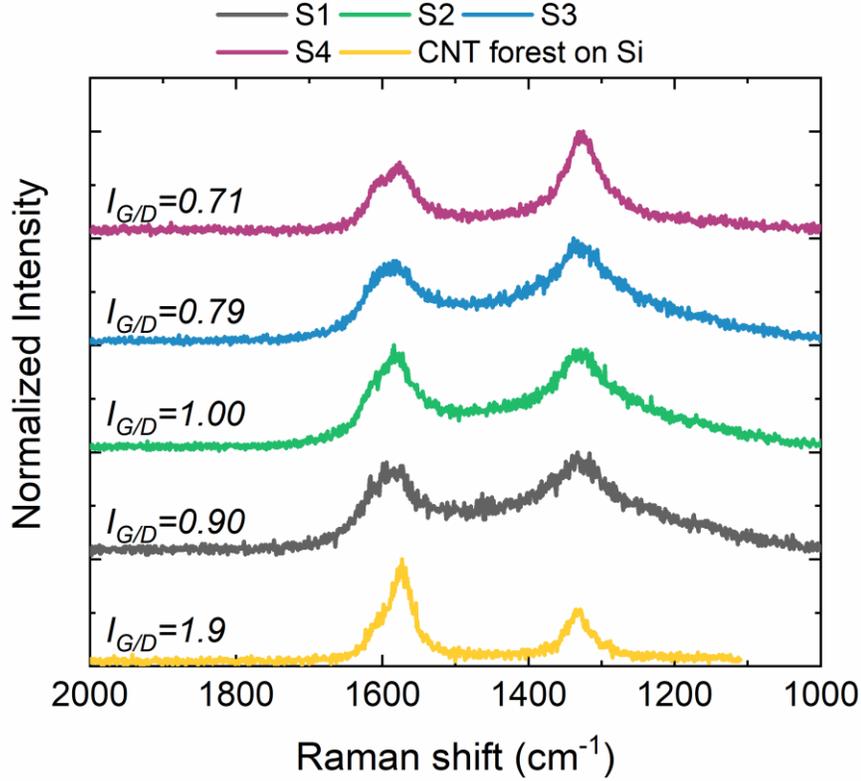

**Figure 2**. Raman spectra obtained from the CNT forests grown on the different substrates. The $I_{G/D}$ ratio is indicated for every substrate type.

### 3.2. Quasi-static characterization

We study the mechanical performance of the CNT forests subjected to quasi-static compressions for five consecutive cycles at a stain rate of 0.02 s$^{-1}$. Results indicate that after the third compression cycle, the forests reach a steady-state response (data for the first five compression cycles for S4 presented in the inset of Fig. 3a). Representative stress-strain curves for the first compression cycle for the different types of steel substrates, as well as for the CNT forest grown on Si wafer are presented in Fig. 3a. From the obtained curves, we extract the dissipated energy (Fig. 3b), unloading modulus (Fig. 3c), peak stress (Fig. 3d), loss coefficient (Fig. 3e) and recovery (Fig. 3f), for each of the five consecutive compression cycles.

At the first compression cycle, samples S1, S2 and S3 have similar dissipated energy values, which range between 1.7-1.9 MJ/m$^3$. The CNT forest on S4 dissipates distinctly less energy, with an average value of 1.3 MJ/m$^3$. CNTs grown on Si dissipate 1.9 MJ/m$^3$, confirming that the energy dissipation is at the same order of magnitude for all CNT forests grown on Si and on steel. For the subsequent cycles, we note a decrease in the amount of dissipated energy, which levels off at about at 20-30% of the original value on all substrates. Same behavior is observed for the CNTs grown on Si wafer.

The peak stress remains almost constant through all the consecutive compressions on all samples, with variations less than ±10%. Specifically, the CNTs on the S1 steel present the highest strength value, peaking at 24.4 MPa, while for S2 the maximum reached stress is 21.2 MPa. For S4 and S3 steels, lower stress levels are observed for the first compression cycle reaching 15.7 and 12.3 MPa, respectively. The peak stress values for the CNTs grown on Si are also constant, with an average at 10 MPa.

The unloading moduli obtained for all samples range between 75 MPa and 100 MPa, with S4 presenting the highest stiffness, followed by S2, S1 and S3. In all steel samples, the unloading modulus is almost constant, with deviations within ±10% for all the successive compression cycles. For CNTs grown on Si, the modulus is found to be higher and constant throughout cyclic loading, averaging at 170 MPa for all cycles. The higher stiffness found on CNTs grown on Si compared to steels, can be correlated to the higher quality and significantly more height of the CNT forests grown on Si.

The loss coefficient, which reflects the degree to which a material dissipates energy, is slightly decreasing for each consecutive cycle for all steel types. We obtain similar values for all samples grown on steels and on Si, which lie between $2 \times 10^{-2}$ and $7 \times 10^{-2}$. We also investigate the deformation recovery of the CNTs after compression. We find that CNTs grown on the steels S1 and S2 show the largest recovery, averaging at 45% for all compressive cycles. The CNTs on S3 and S4 have also an almost constant recovery throughout all cycles, at 23% and 26%, respectively. CNT forests grown on Si recover significantly more, with an average at 78% throughout all compression cycles. This difference for recovery of CNTs grown on steels compared to Si, can be justified by the much higher thickness of the CNT forest on Si, as reported in previous studies [10].

Our results show that the denser, thicker and more aligned CNT forests grown on S1-S2, exhibit the highest energy dissipation, peak stress, loss coefficient and recovery, while the less dense, smaller diameter and more tortuous CNTs grown on sample S4, result in higher stiffness. This is in agreement with previous observations of CNTs grown on Si [10].

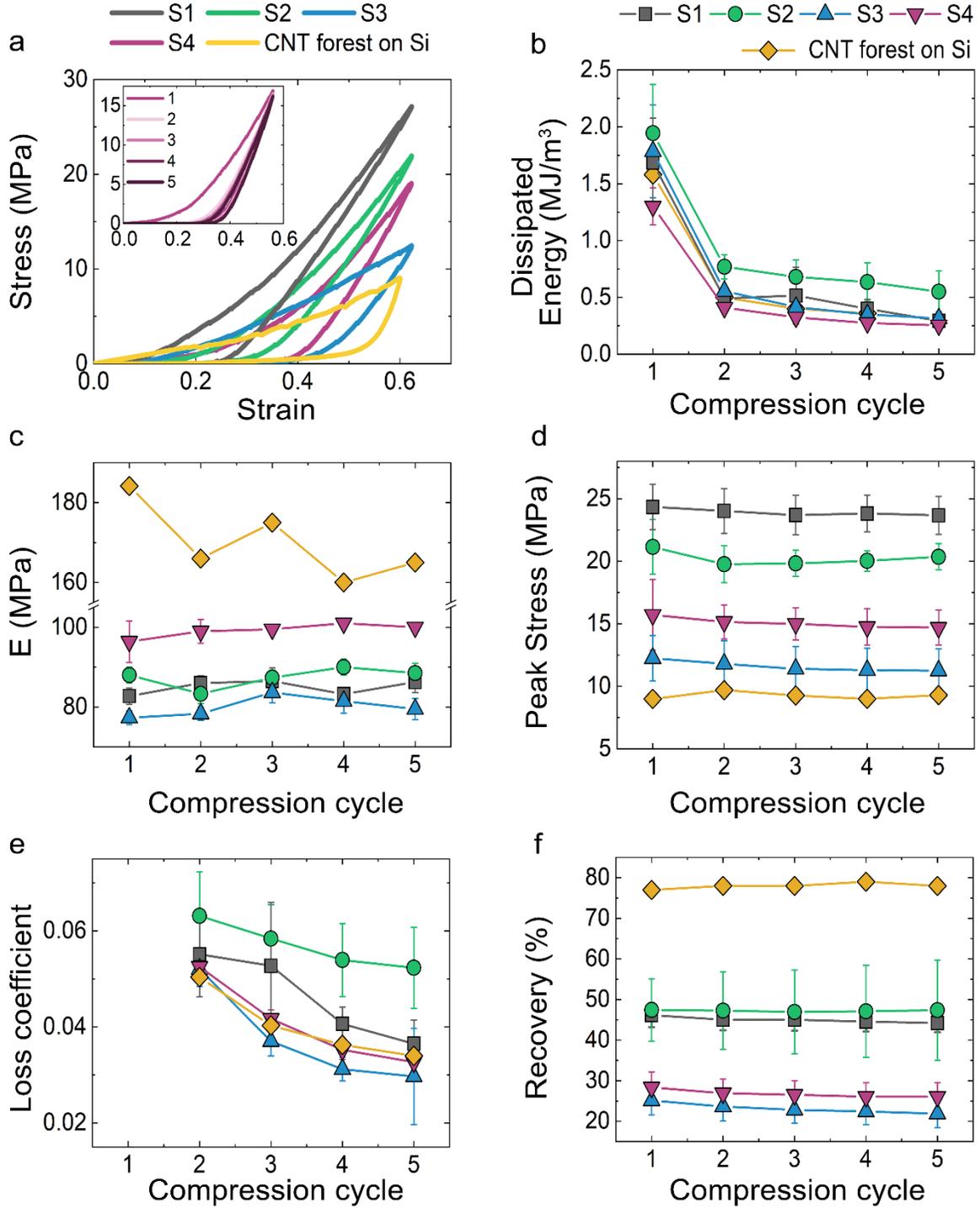

**Figure 3.** (a) Stress-strain curves for CNT forests grown on all substrates; five consecutive compression cycle stress-strain data for S4 (inset); (b) dissipated energy, (c) unloading modulus, (d) peak stress, (e) loss coefficient and (f) recovery percentage for CNT forests grown on all substrates; (b-f) account for five consecutive compression cycles on each sample.

The mechanical performance of the CNT forests suggests that they may be suitable for vibration mitigation and protective applications because they combine high energy dissipation, strength and recoverability. We compare the performance of our materials with other CNT foams or polymeric foams of similar density, which are used in those applications, in an Ashby plot [46] (Fig. 4). The compressive strength of CNT forests grown on metal substrates is superior to the performance of rigid polymer foams and comparable to synthetic materials like high density polyurethanes, polyvinyl chloride and polymethaclylamide (with densities about 1, 0.5 and 0.2 $g/cm^3$ respectively) [46]. However, the rigid polymer foam materials fail catastrophically upon compression, while CNT forests offer the same performance coupled with recoverability. Flexible polymer foams on the other hand, which lie on the bottom left part of the property plot presented in Fig. 4, offer the advantage of recoverability but have considerably lower strength ($10^{-2}$-$10^{0}$ MPa) and energy dissipation ($10^{-3}$-$10^{-1}$ $MJ/m^3$) compared to CNTs. Therefore, the mechanical performance of CNT forests combines advantages otherwise not found in one group of foams.

The Ashby plot of Fig. 4 also offers a comparison with literature-reported mechanical performance of CNT forests. Forests grown on Si fabricated with the floating catalyst method and tested under similar conditions are depicted as grey triangles pointing up and CNT forests fabricated with a different method and/or tested under different conditions are represented by yellow triangles pointing down [1,5,8,9,39,41,64–69]. The comparison confirms our observation that CNT forests grown on steels have similar mechanical properties as CNTs grown on Si.

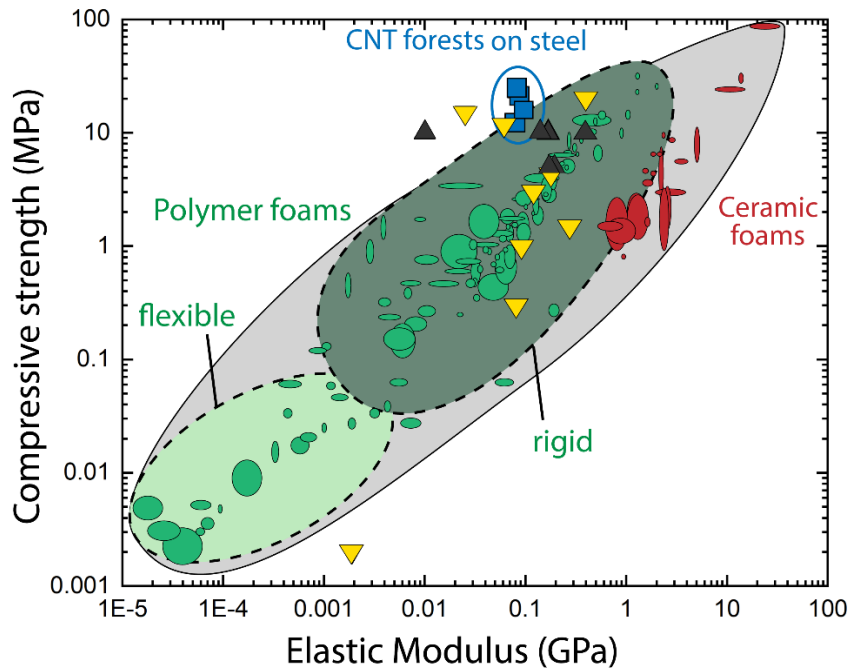

**Figure 4.** Compressive strength versus elastic modulus for foam materials. Green circles represent polymer foams and red circles non-polymer (ceramic and wood) foams. The performance of CNT forests on steel is marked by square, blue markers and CNT forests grown on Si are represented by triangles (pointing up, grey triangles for CNTs fabricated with the same method and tested under similar conditions and pointing down yellow triangles for CNTs fabricated through different methods and/or measured with different conditions). Data from our work and literature-reported values [1,5,8,9,39,41,64–69].

### 3.3. Wetting properties

Corrosion protection on steel surfaces can be obtained using superhydrophobic coatings which prevent water diffusion to the surfaces, thereby extending their service life [21,28,70]. CNTs have been suggested as alternatives because they combine superhydrophobicity with thermal and electrical conductivity, which their organic counterparts lack [21,28,71]. To characterize the hydrophobicity of our coatings, we perform water contact angle measurements. The results are summarized in Table 2.

All the fabricated CNT forests are hydrophobic, and in some cases superhydrophobic. The best performance is obtained for the CNT forest on S2, which has a water contact angle of 161.5º, followed by S3, S1 and S4, for which the contact angles vary from 153.6 to 144.8º. The measured contact angle of the control CNT sample grown on a Si wafer is 133.2 ± 5.8º, which is the lowest measured value. This can be explained by the enhanced surface roughness of the steel surfaces, compared to that of a Si wafer. The higher surface roughness of steels yields a less uniform CNT height, therefore a rougher top surface of the forest. It is known that a higher surface roughness leads to higher contact angles[72,73] and therefore this can count for the higher contact angle of CNT forests on steel versus Si.

The morphological differences of CNTs on the steel substrates can justify the lower hydrophobicity of S4 compared to S1, S2 and S3. As shown by the $I_{G/D}$ index, calculated from the Raman spectra of the samples, S4 presents the highest defect density and tortuosity, and therefore has more non-aligned CNT tips exposed on the surface which justifies the lowest hydrophobicity compared to the rest of the samples. The CNTs on S2 have the highest CNT quality, as indicated by the Raman $I_{G/D}$ index, suggesting more aligned CNT tips on the surface and thus a higher hydrophobicity is expected.

**Table 2.** Contact angle between a water droplet and the CNT forests grown on different substrates.

| Substrate used | Average water contact angle (degrees) |
|---|---|
| S1 | 150.6 ± 10.8 |
| S2 | 161.5 ± 4.4 |
| S3 | 153.6 ± 5.8 |
| S4 | 144.8 ± 0.3 |
| Si | 133.2 ± 5.8º |

### 3.4. Adhesion tests

To evaluate the adhesion between the CNT forests and the steel substrates, we use a custom-made scratching setup (Fig. 5a). As described previously, the force during the razor blade movement along the length of the substrate is continuously monitored. At the time point in which the razor blade scratches the VACNT forest a distinct force increase is recorded, as shown Fig. 5b, in a representative example of an obtained force-time curve, corresponding to sample S3. Using Eq. (6) we calculate the average energy released during scratching for each substrate. The results for all samples, considered at five different scratching speeds, show that the CNTs adhere more efficiently on the surface of S3-S4, compared to S1-S2 at every tested scratch speed (Fig. 5c). Specifically, for S3 the average energy release values range between 530-620 J/m$^2$, while for S4 they are within 250-450 J/m$^2$. For CNTs on S1-S2 the adhesion is almost the same with energy release values around 100-200 J/m$^2$.

As previously discussed and shown in Table 1, the surface energy of steel substrates is highest for S3, and then in descending order for S4, S1 and S2, respectively. Additionally, S3-S4 have larger surface roughness compared to S1-S2. The combination of higher surface energy and roughness, found in S4 and S3, can explain the better adhesion that CNTs have on these surfaces, compared to S1-S2.

Due to the use of different techniques to measure the adhesion-related strength or energy release, a comparison with all literature reported values is not always possible[12,30,51,52]. The results which could be compared with our measurements are those reported by Cao et. al[30], Ageev et. al[52] and Lahiri et. al.[51]. Cao et. grew CNT forests on the end of a silicon carbide (SiC) micron-thick fiber in a form of a brush. They determined the adhesion of the CNTs on SiC by taping over the CNT area and pulling them away from their substrate with a tension machine. The calculated energy based on their results, assuming a pulling length equal to the total length of the bristle in their brush, is 84 J/m$^2$ for CNTs grown on SiC fibers without post-growth treatment[30]. Ageev et. al used the AFM to measure the adhesion of CNT forests grown through PECVD on a Si wafer on a Ti/Ni layer (buffer and catalyst respectively) and reported an adhesion strength of 0.55-1.9 mJ/m$^2$ [52]. Finally, Lahiri et. al, using a nanoindentor scratched through un-oriented CNTs grown on Si and Cu wafers, determined an adhesion of ~80-400 J/m$^2$ for Si and 600-1070 J/m$^2$ for Cu wafers[51]. The results from Lahiri et al. for Si wafer-grown CNTs are comparable to the energy release values found in our work, while those for Cu-grown CNTs indicate a higher adhesion than the one found in our samples. However, it should be noted that in the case of randomly oriented CNTs the contact area between the individual tube and the substrate can be much higher than in the case of vertically-aligned CNTs where mostly only the base of each tube is in contact with the substrate. Therefore, the calculated energy release per unit area can be an overestimation in the case of un-oriented CNTs. The results of Cao et. al, are lower than the values measured in the present work which can be attributed to the different type and geometry of the substrates use in that work. Finally, the results from Ageev et. are significantly lower than the rest of reported values, possibly due to the fundamentally different measurement method employed in that case.

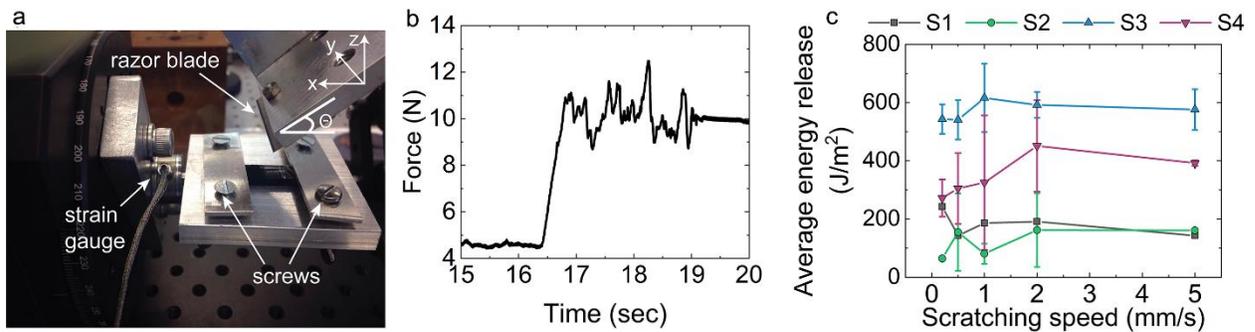

**Figure 5.** (a) Experimental setup for testing the adhesion between the CNTs and their steel substrates. (b) Obtained force-time curve for sample S3. (c) Calculated energy release as a function of the scratching speed, for the different types of samples.

4. Conclusions

In this work, we report the synthesis, structural and mechanical characterization of CNT forests grown on commercial steel substrates. The CNT forests are grown using the floating catalyst method, directly on top of the steel substrates without the use of intermediate adhesive layers, to allow a robust direct contact. We study the mechanical properties, wettability and adhesion of the synthesized CNTs and investigate the structure-property relationships that govern the CNTs' performance. We find that the steel substrate composition and morphology determine the CNT diameter and density. CNTs with larger diameter and density exhibit higher degree of alignment and have the highest energy dissipation, peak stress, loss coefficient and recovery, as well as higher hydrophobicity. CNTs with smaller diameter, density and higher tortuosity result in higher stiffness and lower hydrophobicity. The results reveal that CNT forests grown on steel combine high energy dissipation, high compressive strength and stiffness with recoverability. Their mechanical performance, lightweight and superhydrophobicity, combined with their adhesion to steel, highlight the possibility to use carbon nanotubes forests as a protective multifunctional layer material for metal surfaces.

ASSOCIATED CONTENT

**Supporting Information**.

SEM images of CNTs grown on silicon are presented in Figure S.1

AUTHOR INFORMATION

**Corresponding Author**

* Tel: +1 626 395 8515. E-mail: daraio@caltech.edu (Chiara Daraio)

ACKNOWLEDGMENT

This work was part of the "Advanced cnt structures for functional surfaces applications" project co-funded by ALSTOM/GE and Hightech Zentrum Aargau AG.

Contact angle and surface roughness measurements were performed at the Molecular Materials Research Center of the Beckman Institute of the California Institute of Technology.

The authors would like to thank Dr. Matthew S. Hunt, for the HRTEM imaging.